\documentclass[final,5p,times,twocolumn]{elsarticle}
\usepackage{amssymb}
\usepackage{amsmath,ragged2e}
\usepackage{color,appendix}
\usepackage{graphicx}
\usepackage{caption}
\usepackage{subcaption}
\usepackage[english]{babel}
\usepackage{bm}
\usepackage[percent]{overpic}
\usepackage{multirow,rotating}
\usepackage[english]{babel}

\newcommand{\er}{$\pm$}
\newcommand{\jpsi}{J/\psi}
\newcommand{\beq}{\begin{eqnarray}}
\newcommand{\eeq}{\end{eqnarray}}
\newcommand{\be}{\begin{eqnarray}}
\newcommand{\ee}{\end{eqnarray}}
\newcommand{\bea}{\begin{eqnarray}}
\newcommand{\eea}{\end{eqnarray}}

\newcommand{\nn}{\nonumber\\}

\newcommand{\tz}{$\to$}

\begin{document}
\begin{frontmatter}

\bibliographystyle{try}
\topmargin 0.1cm

\title{Scalar isoscalar mesons and the scalar glueball from radiative $J/\psi$ decays}

\author[label1,label2]{A.V. Sarantsev}
\author[label3]{I. Denisenko}
\author[label1]{U. Thoma}
\author[label1]{and E. Klempt}

\address[label1]{Helmholtz--Institut f\"ur Strahlen-- und Kernphysik, Universit\"at Bonn, Germany}
\address[label2]{NRC ``Kurchatov Institute'', PNPI, Gatchina 188300, Russia}
\address[label3]{Joint Institute for Nuclear Research, Joliot-Curie 6, 141980 Dubna, Moscow region, Russia}

\date{\today}
\begin{abstract}
A coupled-channel analysis of BESIII data on radiative $J/\psi$ decays into $\pi\pi$, $K\bar K$,
$\eta\eta$ and $\omega\phi$ has been performed. The partial-wave amplitude is constrained by a
large number of further data. The analysis finds ten isoscalar scalar mesons. Their masses, widths
and decay modes are determined. The scalar mesons are interpreted as mainly SU(3)-singlet and
mainly octet states. Octet isoscalar scalar states are observed with significant yields only in the
1500-2100\,MeV mass region. Singlet scalar mesons are produced over a wide mass range but their
yield peaks in the same mass region. The peak is interpreted as scalar glueball. Its mass and width are determined to
$M=1865$\er25$^{+10}_{-30}$ {\rm MeV} and $\Gamma= 370$\er$50^{+30}_{-20}$ {\rm MeV}, its yield
in radiative $J/\psi$ decays to ($5.8\pm 1.0)\,10^{-3}$.
\end{abstract}


\end{frontmatter}

\section{Introduction}
Scalar mesons -- mesons with the quantum numbers of the vacuum -- are most fascinating objects in
the field of strong interactions. The lowest-mass scalar meson $f_0(500)$, traditionally often
called $\sigma$, reflects the symmetry breaking of strong interactions and plays the role of the
Higgs particle in quantum chromodynamics (QCD)~\cite{Nambu:1960xd,Nambu:2009zza}. The $f_0(500)$ is
accompanied by further low-mass scalar mesons filling a nonet of particles with spin $J=0$ and
parity $P=+1$: The three charge states $a_0(980)$, the four $K^*_0(700)$, and the two isoscalar
mesons $f_0(980)$, $f_0(500)$ are supposed to be {\it dynamically generated} from meson-meson
interactions~\cite{Pelaez:2003dy}. Alternatively - or complementary - these mesons are interpreted
as four-quark or tetraquark states~\cite{Jaffe:1976ig}.

The continued quest for scalar isoscalar mesons at higher masses  is driven by a
prediction -- phrased for the first time nearly 50 years ago~\cite{Fritzsch:1972jv,Fritzsch:1975tx} -- that QCD
allows for the existence of quark-less particles called glueballs. Their existence is a direct
consequence of the nonabelian nature of QCD and of confinement. However, the strength of the strong
interaction in the confinement region forbids analytical solutions of full QCD. First quantitative
estimates of glueball masses were given in a bag model \cite{DeGrand:1975cf}. Closer to QCD are
calculations on a lattice. In quenched approximation, i.e. when $q\bar q$ loops are neglected, the
lowest-mass glueball is predicted to have scalar quantum numbers, and to have a mass in the 1500 to
1800\,MeV range~\cite{Bali:1993fb,Morningstar:1999rf,Athenodorou:2020ani}; 
unquenching lattice QCD predicts a scalar
glueball at ($1795\pm 60)$\,MeV~\cite{Gregory:2012hu}. Exploiting a QCD Hamiltonian in Coulomb
gauge generating an instantaneous interaction, Szczepaniak and Swanson~\cite{Szczepaniak:2003mr}
calculate the low-lying glueball masses with no free parameters. The scalar glueball of lowest mass
is found at $1980$\,MeV. Huber, Fischer and Sanchis-Alepuz~\cite{Huber:2020ngt} calculate the
glueball spectrum using a parameter-free fully self-contained truncation of Dyson-Schwinger and
Bethe-Salpeter equations and determine the lowest-mass scalar glueball to ($1850\pm 130$)\,MeV.
In gravitational (string) theories -- an analytic approach to QCD --  glueballs are predicted as 
well~\cite{Rinaldi:2021dxh} at 1920\,MeV. Glueballs are predicted consistently within 
a variety of approaches to QCD. They seem to be a safe prediction.

Scalar glueballs are embedded into the spectrum of scalar isoscalar mesons. These have isospin
$I=0$, positive $G$-parity (decaying into an even number of pions), their total spin $J$ vanishes,
their parity $P$ and their C-parity are positive: $(I^G)J^{PC}=(0^+)0^{++}$. Scalar glueballs have
the same quantum numbers as scalar isoscalar mesons and may mix with them. In quark models, mesons
are described as bound states of a quark and an antiquark. Their quantum numbers are often defined
in spectroscopic notation by the orbital angular momentum of the quark and the antiquark $L$, the
total quark spin $S$, and the total angular momentum $J$.  Scalar mesons have
$^{2S+1} L_J=\,^3\hspace{-0.4mm}P_0$.

Experimentally, the scalar glueball was searched for intensively but no generally accepted view has
emerged. The most promising reaction to search for glueballs are radiative decays of $J/\psi$. In
this process, the dominant contribution to direct photon production is expected to come from
the process $J/\psi\to \gamma$ plus two gluons, where the final-state hadrons are 
produced by the hadronization of the two gluons. QCD predicts the two gluons to interact
forming glueballs -- if they exist.  Lattice gauge calculations predict a branching ratio
for radiative $J/\psi$ decays to produce the scalar glueball of ($3.8\pm0.9$)$10^{-3}$
\cite{Gui:2012gx}. This is a significant fraction of all radiative $J/\psi$ decays, (8.8\er1.1)\%. There was hence great excitement when a broad bump in the radiatively produced $\eta\eta$
mass spectrum~\cite{Edwards:1981ex} was discovered by the Crystal Ball collaboration at the
Stanford Linear Accelerator (even though with tensor quantum numbers). However, a resonance 
with the reported properties was not reproduced by any other experiment. The DM2 collaboration
reported a strong peak at 1710\,MeV in the $K\bar K$ invariant mass 
distribution~\cite{Augustin:1987fa}, a peak that is now known as $f_0(1710)$. 
Data  from the CLEO experiment on radiative $J/\psi$ decays into pairs of pseudoscalar mesons were studied 
in a search for glueballs \cite{Dobbs:2015dwa} but no definite conclusions were obtained.

The data with the highest statistics
nowadays stem from BESIII in Bejing. The partial wave amplitudes for  $J/\psi$ radiative decays
into $\pi^0\pi^0$ \cite{Ablikim:2015umt} and $K_SK_S$ \cite{Ablikim:2018izx} were determined in fits to
the data in slices in the invariant mass of the two outgoing mesons. Data on $J/\psi\to
\gamma\eta\eta$ \cite{Ablikim:2013hq} and $J/\psi\to\gamma\phi\omega$~\cite{Ablikim:2012ft} were
presented including an interpretation within a partial wave analysis. In the reactions
$J/\psi\to\gamma 2\pi^+2\pi^-$~\cite{Bai:1999mm,Bugg:2009ch} and
$J/\psi\to\gamma\omega\omega$~\cite{Ablikim:2006ca}, the 
 $2\pi^+2\pi^-$ and into $\omega\omega$ branching ratios of contributing 
resonances were deduced from a smaller data sample.

A new understanding of the spectrum of light-quark scalar mesons emerged from the results obtained
with the Crystal Barrel experiment at the Low-Energy Antiproton Ring at CERN. In $\bar p p$
annihilation at rest, annihilation into $3\pi^0$~\cite{Amsler:1995gf},
$\pi^0\eta\eta$~\cite{Amsler:1995bz}, $\pi^0\eta\eta'$~\cite{Amsler:1994ah}, and
$\pi^0K_LK_L$~\cite{Abele:1996nn} was studied. These data established the existence of the
$f_0(1500)$ resonance; the existence of the $f_0(1370)$ had been proposed in
1966~\cite{Bettini:1966zz} but its existence was accepted only after its rediscovery at LEAR in
$\bar pp$~\cite{Amsler:1994rv} and $\bar pn$ annihilation~\cite{Abele:2001js,Abele:2001pv}.

Central production in hadron-hadron collisions is most\-ly interpreted as collision of two Pomerons,
and this process is supposed to be gluon-rich. Data on this reaction were taken at CERN by the
WA102 collaboration that reported results on  $\pi^+\pi^-$ and $ K_SK_S$ \cite{Barberis:1999cq},
$\eta\eta$ \cite{Barberis:2000cd}, $\eta\eta^{\prime}$ and $\eta^{\prime}\eta^{\prime}$
\cite{Barberis:1999id}, and into  four pions \cite{Barberis:2000em}. The GAMS collaboration
reported a study of the $\pi^0\pi^0$ system in the charge-exchange reactions $\pi^-p\to
\pi^0\pi^0\,n, \eta\eta\,n$ and $\eta\eta'\,n$ at 100\,GeV/c~\cite{Alde:1998mc} 
in a mass range up to 3\,GeV. The charge exchange
reaction $\pi^-p\to K_SK_S\,n$ was studied at the Brookhaven National
Laboratory~\cite{Longacre:1986fh}. An energy-depen\-dent partial-wave analysis based on a slightly
increased data set was reported in Ref.~\cite{Lindenbaum:1991tq}. A reference for any analysis in
light-meson spectroscopy are the amplitudes for $\pi\pi\to\pi\pi$ elastic
scattering~\cite{Grayer:1974cr}. The low-mass $\pi\pi$ interactions are known precisely from the
$K_{\rm e4}$ of charged kaons~\cite{Batley:2010zza}. In these experiments, a series of scalar
isoscalar mesons was found. The Review of Particle Properties (RPP)~\cite{Zyla:2020zbs} lists nine
states; only the five states at lower mass are considered to be established. None of these states
sticks out and identifies itself as the scalar glueball of lowest mass.

Amsler and Close~\cite{Amsler:1995tu,Amsler:1995td} suggested that the three scalar resonances
$f_0(1370)$, $f_0(1500)$ and $f_0(1710)$ could originate from a mixing of the scalar glueball with
a scalar $\frac{1}{\sqrt2}(u\bar u+d\bar d)$ and a scalar $s\bar s$ state. A large number of
follow-up papers suggested different mixing schemes based on these three mesons. We mention here
one recent paper~\cite{Guo:2020akt} that takes into account the production characteristics of these
three mesons in radiative $J/\psi$ decays. However, doubts arose whether a comparatively narrow
$f_0(1370)$ exists \cite{Klempt:2007cp,Ochs:2013vxa}. Klempt and Zaitsev~\cite{Klempt:2007cp}
suggested that the resonances $f_0(1500)$, $f_0(1710)$, $f_0(2100)$ could be SU(3) octet states
while the singlet states merge to a continuous scalar background. This continuous scalar background
was interpreted as scalar glueball by Minkowski and Ochs (and called red
dragon)~\cite{Minkowski:1998mf}. In Refs.~\cite{Anisovich:1996zj,Anisovich:1997ye}, a fourth
isoscalar scalar meson was identified at $1530^{+\ 90}_{-250}$\,MeV with $560\pm 140$\,MeV width.
This broad state was interpreted as glueball. Bugg, Peardon and Zou~\cite{Bugg:2000zy} suggested
that the four known mesons $f_0(1500)$, $f_2(1980)$, $f_0(2105)$, $\eta(2190)$ should be interpreted as
scalar, tensor, excited scalar and pseudoscalar glueball. Recent reviews of scalar mesons and the scalar
glueball can be found elsewhere \cite{Klempt:2007cp,Mathieu:2008me,Crede:2008vw,Ochs:2013gi,Llanes-Estrada:2021evz}.

\section{Our data base}
It seems obvious that the scalar glueball can be identified reliably only once the spectrum of
scalar mesons is understood into which the glueball is embedded. Decisive for the interpretation
are the data on radiative $J/\psi$ decays. But many experiments contribute to our knowledge on
scalar isoscalar mesons and provide additional constraints. In this coupled-channel analysis we fit
meson-pairs in $S$-wave from radiative $J/\psi$ decays and include the $S$-wave contributions to
$\pi\pi$ elastic scattering~\cite{Alde:1998mc} and $\pi\pi\to
K_SK_S$~\cite{Longacre:1986fh,Lindenbaum:1991tq}, the CERN-Munich~\cite{Grayer:1974cr} data and the
$K_{\rm e4}$~\cite{Batley:2010zza} data. Further, we use 15 Dalitz plots for different reactions
from $\bar pN$ annihilation at rest
\cite{Amsler:1995gf,Amsler:1995bz,Abele:1996nn}, \cite{Amsler:2003bq}-\nocite{Abele:1999tf,%
Amsler:1994pz,Abele:1997qy,Wittmack:diss,Abele:1999en}\cite{Abele:1998qd,Abele:1999ac,Abele:1999fw}.

The real and imaginary parts of the mass-dependent $S$-wave amplitudes were derived for $J/\psi\to
\gamma \pi^0\pi^0$  in Ref.~\cite{Ablikim:2015umt} and $J/\psi\to \gamma K_{S} K_{S}$ in
Ref.~\cite{Ablikim:2018izx}. Assuming dominance of resonances with spin $J=0$ and $J=2$, the
partial-wave analysis returned - for each mass bin - two possible solutions, called black ($b$) and red ($r$).
In some mass regions, the two amplitudes practically coincide. We assume continuity between regions
in which the two amplitudes are similar, and divide the full mass range into five regions: in three
regions, the two amplitudes are identical, in two regions, the red and black amplitudes are
different. Thus there are four sets of amplitudes, $(r,r); (r,b); (b,r); (b,b)$. For the data on
$J/\psi\to\gamma K_{S} K_{S}$, we again define five mass regions and four sets of amplitudes. The 
amplitudes ($r,r$) give the best $\chi^2$ for $\pi\pi$, and ($b,b$) for $K_{S} K_{S}$.

\begin{figure*}[pt]
\vspace{-40mm}
\begin{center}
\begin{overpic}[scale=1.35,,tics=10]{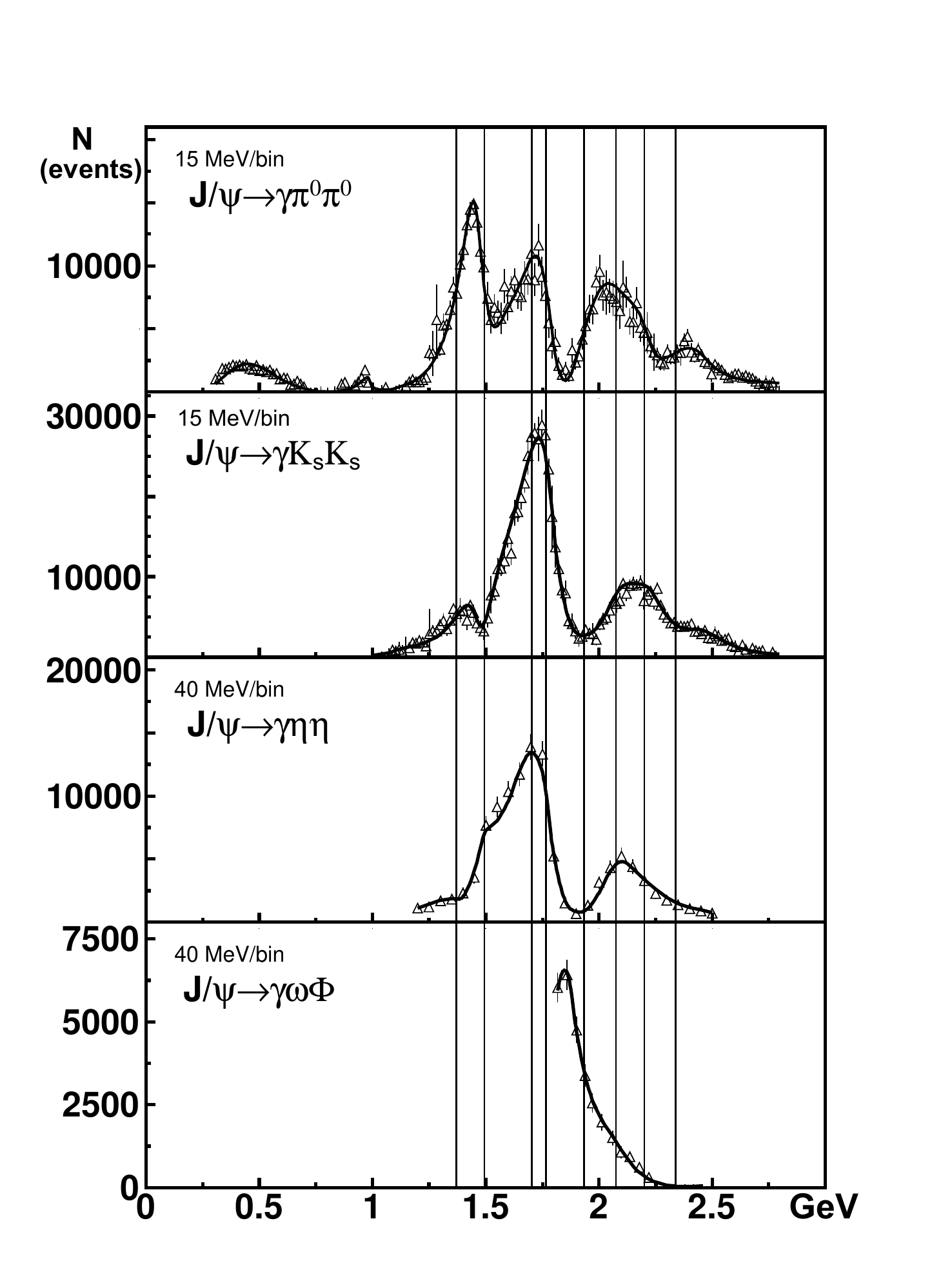}
\put(62,87){\huge\bf a}
\put(62,66){\huge\bf b}
\put(62,45){\huge\bf c}
\put(62,24){\huge\bf d}
\end{overpic}
\vspace{-10mm}
\end{center}
\caption{\label{jpsi}Number of events in the $S$-wave as functions of
the two-meson invariant mass from the reactions
$\jpsi\to \gamma$ $\pi^0\pi^0$ (a), $K_SK_S$~(b), $\eta\eta$ (c),
$\phi\omega$ (d).  (a) and (b) are based on the analysis of 
$1.3\cdot 10^9$ $J/\psi$ decays, (c) and (d) on $0.225\cdot 10^9$ $J/\psi$ decays.
}
\end{figure*}
\begin{figure}[pt]
\vspace{-10mm}
\begin{center}
\begin{overpic}[scale=0.6,tics=10]{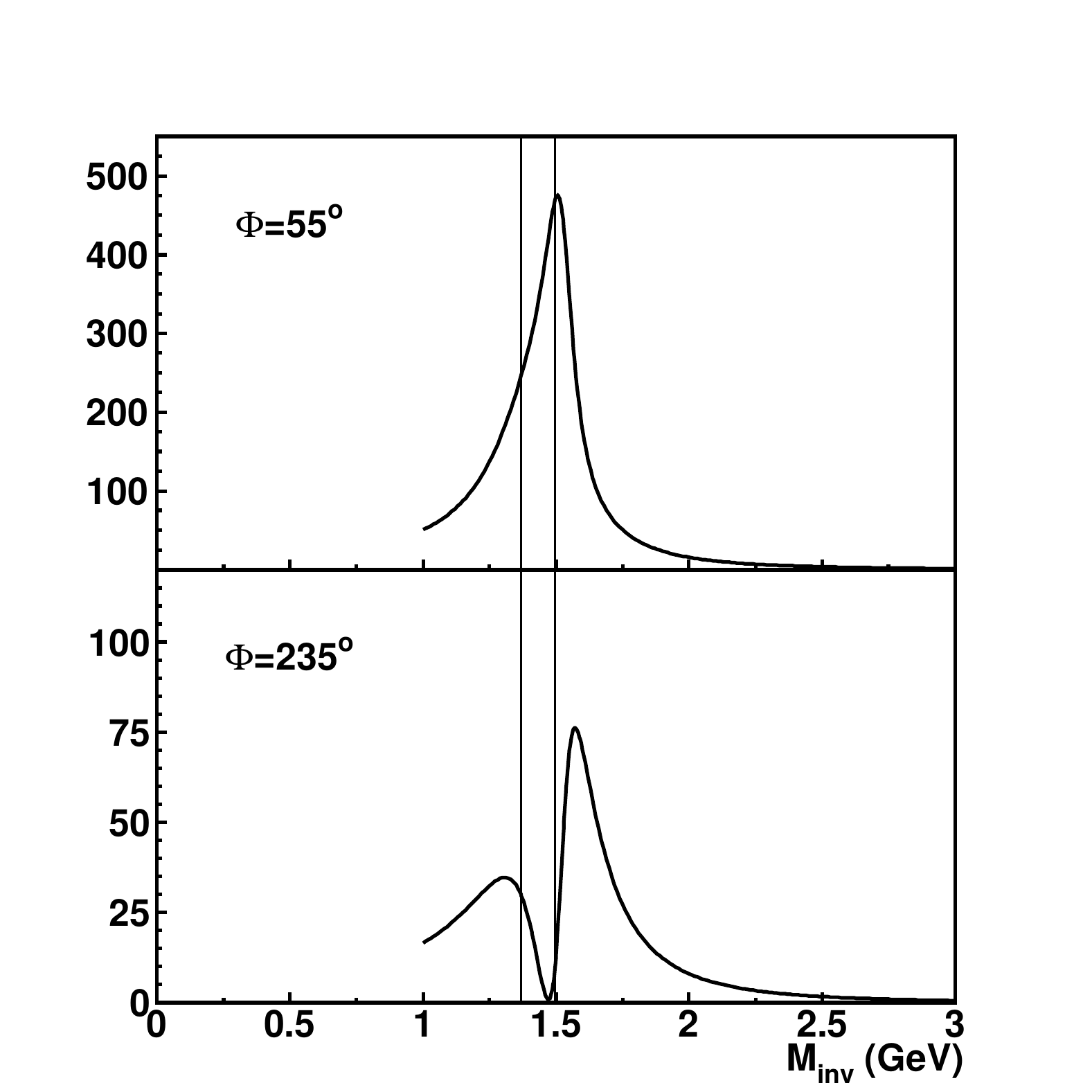}
\put(81,82){\huge a}
\put(81,40){\huge b}
\end{overpic}
\vspace{-6mm}
\end{center}
\caption{\label{jpsi2}Interference of two $K$-matrix poles for $f_0(1370)$ and $f_0(1500)$
with a relative phase of $55^\circ$ (a) and  $235^\circ$ (b). Shown is the intensity in arbitrary units
as a function of the two-meson invariant mass.
}
\end{figure}
Figure~\ref{jpsi}a,b shows the  $\pi^0\pi^0$~\cite{Ablikim:2015umt} and
$K_SK_S$~\cite{Ablikim:2018izx} invariant mass distributions from radiative $J/\psi$ decays for the
best set of amplitudes. The ``data'' are represented by triangles with error bars, the solid curve
represents our fit. (i) The $\pi^0\pi^0$ invariant mass distribution shows rich structures. So far, no
attempt has been reported to understand the data within an energy-dependent partial-wave analysis.
The mass distribution starts with a wide enhancement at about 500\,MeV and a narrow peak at
975\,MeV: with $f_0(500)$ and $f_0(980)$. A strong enhancement  follows peaking at 1450\,MeV, and a
second minimum at 1535\,MeV. Three more maxima at 1710\,MeV, 2000\,MeV, and 2400\,MeV are separated by a deep minimum at about 1850 MeV and a weak one at 2285\,MeV. 
(ii) The $K_SK_S$ invariant mass
distribution exhibits a small peak at 1440\,MeV immediately followed by a sharp dip at 1500\,MeV.
The subsequent enhancement - with a peak position of 1730\,MeV - has a significant low-mass
shoulder. There is a wide minimum at 1940\,MeV followed by a further structure peaking at
2160\,MeV. It is followed by a valley at 2360\,MeV, a small shoulder above
2400\,MeV, and a smooth continuation. (iii) Data on $J/\psi\to \gamma\eta\eta$ \cite{Ablikim:2013hq} and
$J/\psi\to\gamma\phi\omega$~\cite{Ablikim:2012ft} were published including an interpretation within a partial
wave analysis. We extracted the $S$-wave contributions (Fig.~\ref{jpsi}c,d) (for ~\cite{Ablikim:2012ft} only
the contribution of the dominant threshold resonance), and included them in the fits. The $\eta\eta$ mass distribution
(Fig.~\ref{jpsi}d) resembles the one observed at SLAC~\cite{Edwards:1981ex}, but now greatly improved
statistics. The BESIII data show an asymmetry that reveals the presence of at least two resonances,
$f_0(1500)$ and $f_0(1710)$. The sharp drop-off above 1.75\,GeV indicates destructive interference
between two resonances. (iv) The
$\phi\omega$ mass distribution exhibits a strong threshold enhancement. It was
assigned to a scalar resonance at (1795\er 7) MeV~\cite{Ablikim:2012ft} but the data can also be
described by a resonance at 1770\,MeV that was suggested earlier~\cite{Bugg:2004xu}
and that is required in our fit.

The two figures \ref{jpsi}a,b look very different. Obviously, interference between neighboring
states plays a decisive role. Figures~\ref{jpsi2}a,b simulate the observed pattern: in
Fig.~\ref{jpsi2}a the low-mass part of $f_0(1500)$ interferes constructively with $f_0(1370)$ and leads to a 
sharp drop-off at its high-mass part. In
Fig.~\ref{jpsi2}b, the $f_0(1500)$ is responsible for the dip. The phase difference between the
amplitudes for $f_0(1370)$ and $f_0(1500)$ in figures \ref{jpsi2}a,b changes by $180^\circ$ when going from $\pi\pi$ to
$K\bar K$. This is an important observation: these two states do not behave like a $\frac{1}{\sqrt
2}(u\bar u+d\bar d)$ and a $s\bar s$ state but rather like a singlet and an octet state, like 
$\frac{1}{\sqrt 3}(u\bar u+d\bar d+s\bar s)$ and  $\frac{1}{\sqrt 6}(u\bar u+d\bar d -2s\bar s)$.
This change in the sign of the coupling constant in $\pi\pi$ and $K\bar K$ decays for $f_0(1500)$ with respect to
the $f_0(1370)$ ``background'' has first been noticed in Ref.~\cite{Minkowski:2004xf}.

The two resonances $f_0(1710)$ and $f_0(1770)$ form the large enhancement in figure~\ref{jpsi}b
while their contribution to figure~\ref{jpsi}a is much smaller. This again is due to interference:
the two resonances interfere destructively in the $\pi\pi$ channel and constructively in the $K\bar K$
channel. Again, the $f_0(1710)$ and $f_0(1770)$ wave functions must contain significant
$u\bar u+d\bar d$ and $s\bar s$ contributions of opposite signs: there must one singlet-like
and one octet-like state.

In $J/\psi$ radiative decays, the final-state mesons are produced by two gluons in the initial state.
It is illuminating to compare the  $\pi\pi$ and $K\bar K$ mass distributions shown
in Figs.~\ref{jpsi} with the ones produced when an 
$s\bar s$ pair forms the initial state. 
 Ropertz, Hanhart and Kubis~\cite{Ropertz:2018stk} 
analyzed the $\pi\pi$ and $K\bar K$ 
systems produced in the reaction $\overline{B}^0_s\to J/\psi \pi^+\pi^-$ \cite{Aaij:2014emv} and 
$\overline{B}^0_s\to J/\psi K^+K^-$ \cite{Aaij:2017zgz}. Here, the $\pi\pi$ and $K\bar K$ systems 
stem from an $s\bar s$ pair recoiling against the $J/\psi$. 
The pion and the kaon form factors are dominated by $f_0(980)$, followed by a bump-drop-off 
($\pi\pi$) or peak ($K\bar K$) structure at 1500\,MeV and a small enhancement just below 
2000\,MeV. The form factors (as well as the mass spectra) are 
decisively different from the spectra shown in Fig.~\ref{jpsi}
that originate from two interacting gluons and that are dominated by a large intensity in the
1700 to 2100\,MeV mass range.

\section{The PWA}

The different sets of partial-wave amplitudes were fitted with a modified $K$-matrix 
approach~\cite{Anisovich:2011zz} that takes into account dispersive corrections and the Adler zero. 
We fit data in which resonances are produced and data in which they are
formed in scattering processes. 
The scattering amplitude between the channels $a$ and $b$ is
described as
\be
A_{ab}=\sum\limits_{\alpha,\beta} g^{R(\alpha )}_a d_{\alpha\alpha}
D_{\alpha\beta}g^{L(\beta)}_b\,,
\label{ampl}
\ee
while the production of a resonance is given by a P-vector amplitude:
\beq
\hspace{-6mm}A_{b}\!=\!\sum\limits_{\alpha,\beta} \tilde P^{(\alpha )}
d_{\alpha\alpha} D_{\alpha\beta}g^{L(\beta)}_b~~\tilde P\!=\!\left
(\Lambda_1,\ldots\Lambda_n,F_1,\ldots\right )\,.
\label{P-vector}
\eeq
The $\Lambda_\alpha$ are production couplings of the resonances,
the $F_j$ represent non-resonant transitions from the initial to
the final states, and the $g^{R(\alpha )}_a$ and $g^{L(\alpha)}_b$ are right-hand ($R$)
and left-hand ($L$) coupling constants of the state $\alpha$ into channels
$a$ and $b$. Here the ``state"  represents either the bare resonant
state or a non-resonant contribution. For resonances
the vectors of right-hand and left-hand vertices are identical (but transposed), for 
non-resonant contributions, the vertices can be different, even the sign can differ. The
$d_{\alpha\alpha}$ are elements of the diagonal matrix of the
propagators $\hat d$:
 \be
 \hat d=diag\left
(\frac{1}{M^2_1-s},\ldots,\frac{1}{M^2_N-s},R_1,R_2 \ldots\right),
\ee
where the first $N$ elements describe propagators of 
resonances and $R_\alpha$ are propagators of non-resonant
contributions. Here, these are constants and or a pole well below the
$\pi\pi$ threshold.

The block $D_{\alpha\beta}$ describes the transition between the
bare state $\alpha$ and the bare state $\beta$ (with the propagator
of the $\beta$ state included). For this block one can write the
equation (with summation over the double indices $\gamma, \eta$)
\be
D_{\alpha\beta}= D_{\alpha\gamma}\sum\limits_j
B^j_{\gamma\eta}d_{\eta\beta}+d_{\alpha\beta} \label{eqbs}.
\ee
The elements $B^j_{\gamma\eta}$ describe the loop diagrams of
the channel $j$ between states $\gamma$ and $\eta$:
\be
B^j_{\alpha\beta}=
\int\frac{ds'}{\pi}\frac{g^{R(\alpha)}_j\rho_j(s',m_{1j},m_{2j})g^{L(\beta)}_j}{s'-s-i0}
\;.
\label{loop}
\ee
In matrix form, Eqn.~(\ref{eqbs}) can be written as
 \be
 \hat D= \hat D\hat B\hat d+\hat d \qquad \hat D= \hat
d(I-\hat B\hat d)^{-1},
 \ee
where the elements of the $B_{\alpha\beta}$ matrix are equal to the sum
of the loop diagrams between states $\alpha$ and $\beta$:
\be
\hat B_{\alpha\beta}=\sum\limits_j B^j_{\alpha\beta}\;.
\ee
If only the imaginary part of the integral from Eqn.~(\ref{loop}) is
taken into account, Eqn.~(\ref{ampl}) 
corresponds to the standard $K$-matrix amplitude. In this case, the energy
dependent non-resonant terms can be described by left-hand
vertices only, the right-hand vertices can be set to 1.

Here, the real part of the loop diagrams is taken into account, and we 
parameterize the non-resonant contributions 
either as constants or as a pole below all relevant thresholds. The latter
parameterization reproduces well the projection of the $t$ and
$u$-channel exchange amplitudes into particular partial waves.

The elements of the
$B^j_{\alpha\beta}$ are calculated using one subtraction taken at the channel
threshold $M_{j}=(m_{1j}+m_{2j})$:
\be
B^{j}_{\alpha\beta}(s)&=&B^j_{\alpha\beta}(M_{j}^2)+(s-M_j^2)\nn&\times&
\int\limits_{M_j^2}^\infty \frac{ds'}{\pi}
\frac{g^{R(\alpha)}_j\rho_j(s',m_{1j},m_{2j})g^{L(\beta)}_j}{(s'-s-i0)(s'-M^2_{j})}.
\label{D9} \ee
Our parameterization of the non-resonant contributions allows us to rewrite 
the $\hat B$-matrix as
\be
B^{j}_{\alpha\beta}(s)=g^{R(\alpha)}_a\left ( b_0^j+ (s-M_j^2)
b_j(s)\right ) g^{L(\beta)}_b\,,
\label{bfull}
\ee
where the parameters $b^j$ depend on decay channels only:
\be
b_j(s)=\int\limits_{M_j^2}^\infty \frac{ds'}{\pi}
\frac{\rho_j(s',m_{1a},m_{2a})}{(s'-s-i0)(s'-M^2_{j})}\,.
\label{bj}
\ee
In this form the $D$-matrix approach would be equivalent to the $K$-matrix
approach when the substitution
\be
i\rho_j(s)\to b_0^j+ (s-M_j^2) b_j(s)
\ee
is made. The  Adler zero (set to $s_A=m_\pi ^2/2$) is introduced by a modification of
the phase volume, with $s_{A_0}$=0.5\,GeV$^2$:
\be
\rho_1(s,m_\pi,m_\pi)=\frac{s-s_A}{s+s_{A_0}}\sqrt{\frac{s-4m^2_\pi}{s}}\,.
\ee
Branching ratios of a resonance into the final state $\alpha$
were determined by defining a Breit-Wigner amplitude in the form
\be
A_\alpha=\frac{f\,g_{J/\psi}\ g_\alpha}{M_0 ^2-s-i\,
f\sum\limits_\alpha g_{\alpha}^2 \ \rho^{\alpha}(s)}
\ee

The Breit-Wigner mass $M_0$ and the parameter $f$ are fitted to reproduce
the $T$-matrix pole position. For all states the factor $f$ was between
0.95 and 1.10; the Breit-Wigner mass exceeded the pole mass by $10-20$\,MeV.
The decay couplings $g^{\alpha}$ and production couplings 
$g_{J/\psi}$ were calculated as residues at the pole
position. Then we use the definition (Eqn.~(49.16) in \cite{Zyla:2020zbs})
\be
\hspace{-8mm}M_{0}\Gamma^{\alpha}  = (g^{\alpha})^2 \ \rho^{\alpha}
({M_{0}}^{2}); &\qquad 
BR^\alpha  = \Gamma^{\alpha}/\Gamma_{\rm BW}.
\label{br}
\ee
Branching ratios into final states with little phase space only were determined by integration
over the (distorted) Breit-Wigner function. This procedure is used in
publications of the Bonn-Gatchina PWA group and was compared to
other definitions in Ref.~\cite{Nikonov:2018pr}.

The $K$-matrix had couplings to $\pi\pi$, $K\bar K$, $\eta\eta$, $\eta\eta'$,
$\phi\omega$, to $\omega\omega$, and to the four-pion phase-space representing unseen multibody
final states. The $\omega\omega$ and the four-pion intensities are treated as {\it missing} intensity.  
Fits with $K$-matrix poles above 1900\,MeV were found to be unstable: the CERN-Munich data on elastic scattering
and the GAMS data on $\pi\pi\to \pi^0\pi^0, \eta\eta$ and $\eta\eta'$ stop at about 1900\,MeV.
For resonances above, only the product of the coupling constants for production and decay can
be determined. Therefore 
we used $K$-matrix poles for resonances below and Breit-Wigner amplitudes above 1900\,MeV.
The latter amplitudes had the form
\be
A_{\rm BW}=\frac{g_{\rm BW}}{M_{\rm BW} ^2-s-i\,M_{\rm BW}\,\Gamma_{\rm BW}}\,.
\label{BW}
\ee
\begin{table}[pt]
\caption{\label{chisquare}$\chi^2/N_{\rm data}$ contribution from radiative $J/\psi$ decays, charge
exchange reactions, $K_{\rm e4}$ decays, and Dalitz plots from $\bar pN$ annihilation at rest.
Dalitz plots are given by the number of events in cells, $N$ is the number of cells. }
\renewcommand{\arraystretch}{1.2}
\centering\small
\begin{tabular}{|lccc||lccc|}
\hline\hline
$J/\Psi\!\to$         \hspace{-5mm}&\hspace{-5mm}  $\chi^2/N$ \hspace{-5mm}&\hspace{-5mm}  $N$ \hspace{-5mm}&\hspace{-5mm} Ref. & $\bar p p(liq)\to$\hspace{-5mm}&\hspace{-5mm}  $\chi^2/N$ \hspace{-3mm}&\hspace{-3mm}  $N$ \hspace{-3mm}&\hspace{-3mm} Ref. \\\hline
$\gamma \pi^0\pi^0$  \hspace{-5mm}&\hspace{-5mm}  1.28   \hspace{-5mm}&\hspace{-5mm}  167  \hspace{-5mm}&\hspace{-5mm} \cite{Ablikim:2015umt}&   $\pi^0\pi^0\pi^0$ \hspace{-5mm}&\hspace{-5mm}  1.40 \hspace{-3mm}&\hspace{-3mm}  7110\hspace{-3mm}&\hspace{-3mm}\cite{Amsler:1995gf} \\
$\gamma K_S K_S$     \hspace{-5mm}&\hspace{-5mm}  1.21   \hspace{-5mm}&\hspace{-5mm}  121  \hspace{-5mm}&\hspace{-5mm} \cite{Ablikim:2018izx}&  $\pi^0\eta\eta$   \hspace{-5mm}&\hspace{-5mm}  1.28 \hspace{-3mm}&\hspace{-3mm} 3595\hspace{-3mm}&\hspace{-3mm}\cite{Amsler:1995bz} \\
$\gamma \eta\eta$      \hspace{-5mm}&\hspace{-5mm}  0.80   \hspace{-5mm}&\hspace{-5mm}  21    \hspace{-5mm}&\hspace{-5mm} \cite{Ablikim:2013hq}&   $\pi^0\pi^0\eta$  \hspace{-5mm}&\hspace{-5mm}  1.23 \hspace{-3mm}&\hspace{-3mm}  3475\hspace{-3mm}&\hspace{-3mm} \cite{Amsler:1994pz} \\
$\gamma \phi\omega$ \hspace{-5mm}&\hspace{-5mm}  0.2      \hspace{-5mm}&\hspace{-5mm}  17    \hspace{-5mm}&\hspace{-5mm} \cite{Ablikim:2012ft}&  $\pi^+\pi^0\pi^-$ \hspace{-5mm}&\hspace{-5mm} 1.24 \hspace{-3mm}&\hspace{-3mm}1334\hspace{-3mm}&\hspace{-3mm} \cite{Abele:1997qy}\\
\cline{1-4}
$\pi^-\pi^+\!\!\to$ \hspace{-5mm}&\hspace{-5mm} \hspace{-5mm}&\hspace{-5mm}  \hspace{-5mm}&\hspace{-5mm}                             &   $K_LK_L\pi^0$ \hspace{-5mm}&\hspace{-5mm}  1.08 \hspace{-3mm}&\hspace{-3mm}  394\hspace{-3mm}&\hspace{-3mm} \cite{Abele:1996nn}\\
\cline{1-4}
$\pi^0\pi^0$         \hspace{-5mm}&\hspace{-5mm}  0.89         \hspace{-5mm}&\hspace{-5mm}  110  \hspace{-5mm}&\hspace{-5mm} \cite{Alde:1998mc}&  $K^+K^-\pi^0$ \hspace{-5mm}&\hspace{-5mm}  0.97 \hspace{-3mm}&\hspace{-3mm}521\hspace{-3mm}&\hspace{-3mm} \cite{Abele:1999en}\\
$\eta\eta$             \hspace{-5mm}&\hspace{-5mm}  0.67         \hspace{-5mm}&\hspace{-5mm}  15    \hspace{-5mm}&\hspace{-5mm} \cite{Alde:1998mc}& $K_SK^\pm\pi^\mp$ \hspace{-5mm}&\hspace{-5mm}  2.13 \hspace{-3mm}&\hspace{-3mm}  771\hspace{-3mm}&\hspace{-3mm} \cite{Wittmack:diss} \\
$\eta\eta'$            \hspace{-5mm}&\hspace{-5mm}  0.23         \hspace{-5mm}&\hspace{-5mm}  9      \hspace{-5mm}&\hspace{-5mm} \cite{Alde:1998mc}&$K_LK^\pm\pi^\mp$ \hspace{-5mm}&\hspace{-5mm}  0.76 \hspace{-3mm}&\hspace{-3mm}  737\hspace{-3mm}&\hspace{-3mm} \cite{Abele:1998qd}\\
\cline{5-8}
$K^+K^-$            \hspace{-5mm}&\hspace{-5mm}  1.06         \hspace{-5mm}&\hspace{-5mm}  35    \hspace{-5mm}&\hspace{-5mm} \cite{Lindenbaum:1991tq}& $\bar p n(liq)\to$    \hspace{-5mm}&\hspace{-5mm}  \hspace{-5mm}&\hspace{-5mm} \hspace{-5mm}&\hspace{-5mm} \\
\cline{5-8}
$\pi^+\pi^-$        \hspace{-5mm}&\hspace{-5mm}  1.32          \hspace{-5mm}&\hspace{-5mm}  845   \hspace{-5mm}&\hspace{-5mm}  \cite{Grayer:1974cr}&   $\pi^+\pi^-\pi^-$\hspace{-5mm}&\hspace{-5mm}  1.39 \hspace{-3mm}&\hspace{-3mm} 823 \hspace{-3mm}&\hspace{-3mm} \cite{Abele:1999tf} \\
$\delta(elastic)$    \hspace{-5mm}&\hspace{-5mm} 0.91           \hspace{-5mm}&\hspace{-5mm}  17     \hspace{-5mm}&\hspace{-5mm} \cite{Batley:2010zza}&  $\pi^0\pi^0\pi^-$\hspace{-5mm}&\hspace{-5mm}  1.57 \hspace{-3mm}&\hspace{-3mm} 825\hspace{-3mm}&\hspace{-3mm} \cite{Abele:1997qy}   \\
\cline{1-4}
$\bar p p(gas)\to$ \hspace{-5mm}&\hspace{-5mm}  \hspace{-5mm}&\hspace{-5mm}  \hspace{-5mm}&\hspace{-5mm}                    &  $K_SK^-\pi^0$    \hspace{-5mm}&\hspace{-5mm}  1.33 \hspace{-3mm}&\hspace{-3mm} 378   \hspace{-3mm}&\hspace{-3mm} \cite{Wittmack:diss}  \\
\cline{1-4}
$\pi^0\pi^0\pi^0$ \hspace{-5mm}&\hspace{-5mm}  1.36          \hspace{-5mm}&\hspace{-5mm} 4891 \hspace{-5mm}&\hspace{-5mm} \cite{Amsler:2003bq}  &   $K_SK_S\pi^-$    \hspace{-5mm}&\hspace{-5mm}  1.62 \hspace{-3mm}&\hspace{-3mm} 396    \hspace{-3mm}&\hspace{-3mm} \cite{Wittmack:diss}    \\
$\pi^0\eta\eta$    \hspace{-5mm}&\hspace{-5mm} 1.32             \hspace{-5mm}&\hspace{-5mm}  1182 \hspace{-5mm}&\hspace{-5mm} \cite{Amsler:2003bq}\hspace{-5mm}&\hspace{-5mm} \hspace{-5mm}&\hspace{-5mm} \hspace{-5mm}&\hspace{-5mm} \hspace{-5mm}&\hspace{-5mm}    \\
$\pi^0\pi^0\eta$  \hspace{-5mm}&\hspace{-5mm} 1.24             \hspace{-5mm}&\hspace{-5mm}  3631 \hspace{-5mm}&\hspace{-5mm} \cite{Abele:1999tf}\hspace{-5mm}&\hspace{-5mm} \hspace{-5mm}&\hspace{-5mm} \hspace{-5mm}&\hspace{-5mm} \hspace{-5mm}&\hspace{-5mm}   \\
\hline\hline
\end{tabular}
\vspace{-4mm}
\end{table}
The total amplitude was thus written as the sum of the $P$-vector
amplitude (Eqn. \ref{P-vector}) and a summation over for Breit-Wigner amplitudes (Eqn.~\ref{BW}).

Table~\ref{chisquare} gives the $\chi^2$ of our best fit for the various data sets. This fit requires 
contributions from ten resonances. Their masses and widths are given in
Table~\ref{MW}, their decay properties in Table~\ref{decays}. The errors stem from the spread of
results from different sets of $S$-wave amplitudes \cite{Ablikim:2015umt,Ablikim:2018izx}, and
cover the spread of results when the background amplitude was altered (constant transition
amplitudes or left-hand pole in the $K$-matrix), when the number of high-mass resonances was changed
(between 7 and 11), and part of the data were excluded from the fit. The resulting
values are compared to values listed in the RPP~\cite{Zyla:2020zbs}. The overall agreement is rather good. Only the five low-mass resonances are classified as
established in the RPP, the other states needed confirmation. We emphasize that the solution
presented here was developed step by step, independent from the RPP results. The comparison was
made only when the manuscript was drafted.

The sum of Breit-Wigner amplitudes is not manifestly unitary. However, radiative $J/\psi$ decays
and the two-body decays of high-mass resonances are far from the unitarity limit. To check possible 
systematic errors due to the use of Breit-Wigner amplitudes, we replace the four resonances
$f_0(1370)$, $f_0(1500)$, $f_0(1710)$, $f_0(1770)$ by Breit-Wigner amplitudes (imposing
mass and width of $f_0(1370)$). The fit returns properties of these resonances within the errors
quoted in Table \ref{MW} and~\ref{decays}.

The four lower-mass resonances,  one scalar state at about 1750\,MeV and the $f_0(2100)$, are
mandatory for the fit: if one of them is excluded, no acceptable description of the data is
obtained. Only one state, $f_0(1770)$, is ``new''.  Based on different peak positions,
Bugg~\cite{Bugg:2004xu} had suggested that $f_0(1710)$ should have a close-by state called
$f_0(1770)$. When $f_0(1710)$ and $f_0(1770)$ are replaced by one resonance, the $\chi^2/N_{\rm
data}$ increases by 58/167 for $J/\psi\to\gamma\pi^0\pi^0$, 8/121 for $J/\psi\to\gamma K_SK_S$, 50/21
for $J/\psi\to\gamma\eta\eta$, or by (58, 8, 50) in short. When $f_0(2020)$, $f_0(2200)$, or
$f_0(2330)$ are removed, the $\chi^2$ increases by (48, 6, 5); (30,6,1); (23, 5, 0). In addition,
there is a very significant deterioration of the fit to the Dalitz plots for $\bar pp$ annihilation when
only one scalar resonance in the 1700 to 1800\,MeV range is admitted. When
high-mass poles are removed, a small change in $\chi^2$ is observed also in the data on
$\bar pp$ annihilation due to a change of the interference
between neighboring poles. All ten states contribute to the reactions studied here.

\section{The flavor wave functions}
The interference pattern in Fig.~\ref{jpsi}a,b suggests that the
two pairs of resonances, $f_0(1370)$-$f_0(1500)$ and $f_0(1710)$ -$f_0(1770)$, have wave flavor functions 
in which the $u\bar u+d\bar d$ and $s\bar s$ components have opposite signs. 
Vector ($\omega, \phi$) and tensor ($f_2(1270), f_2'(1525)$) mesons show {\it ideal} mixing, with
approximately $1/\sqrt{2}(u\bar u+d\bar d)$ and $s\bar s$ configurations. (We
neglect the small mixing angles in this discussion.) The mass difference is 200 to 250\,MeV, and
the $s\bar s$ mesons decay preferably into $K\bar K$. This is not the case for the scalar mesons:
The mass difference varies, and the $K\bar K / \pi\pi$ decay ratio is often less than 
1 and never $\sim100$ like for $f_2'(1525)$. The decay pattern rules 
out the possibility that the scalar resonances are states with a dominant
$s\bar s$ component.

Pseudoscalar mesons are different: the isoscalar mesons are better
approximated by SU(3) singlet and octet configurations. Pseudoscalar mesons have both,
$1/\sqrt{2}(u\bar u+d\bar d)$ and $s\bar s$ components as evidenced by the comparable rates for
$J/\psi \to \eta\omega, \eta\phi, \eta^\prime\omega , \eta^\prime\phi$.

In view of these arguments and the interference pattern discussed above, we make a very simple
assumption: we assume that the upper states in Table~\ref{MW} all have large SU(3)-singlet components, 
while the lower states have large octet components. For the two lowest-mass mesons, $f_0(500)$
and $f_0(980)$, Oller~\cite{Oller:2003vf} determined the mixing angle to be small, 
(19\er5)$^\circ$: $f_0(500)$ is dominantly SU(3) singlet, $f_0(980)$ mainly octet.

We choose $f_0(1500)$ as reference state and plot a $(M^2, n)$ trajectory with $M_n
^2=1.483^2 + n\,a$\,GeV$^2, n=-1, 0, 1, \cdots$, where $a=1.08$ is the slope of the 
trajectory. States close to this trajectory are assumed to be mainly SU(3) octet states. In
instanton-induced interactions, the separation in mass square of scalar singlet and octet mesons is
the same as the one for pseudoscalar mesons, but reversed~\cite{Klempt:1995ku}. Hence we calculate
a second trajectory $m_n^2= 1.483^2 + m_{\eta}^2 - m_{\eta'}^2  + n\,a$\,GeV$^2, n=-1, 0, 1,
\cdots$. The low-mass singlet mesons are considerably wider than their octet partners.  With
increasing mass, the width of singlet mesons become smaller (except for $f_0(2020)$), those of
octet mesons increase (except for $f_0(2330)$).

\begin{figure}[pt]\vspace{-5mm}
\hspace{-2mm}\includegraphics[width=0.50\textwidth,height=0.34\textwidth,clip=on]{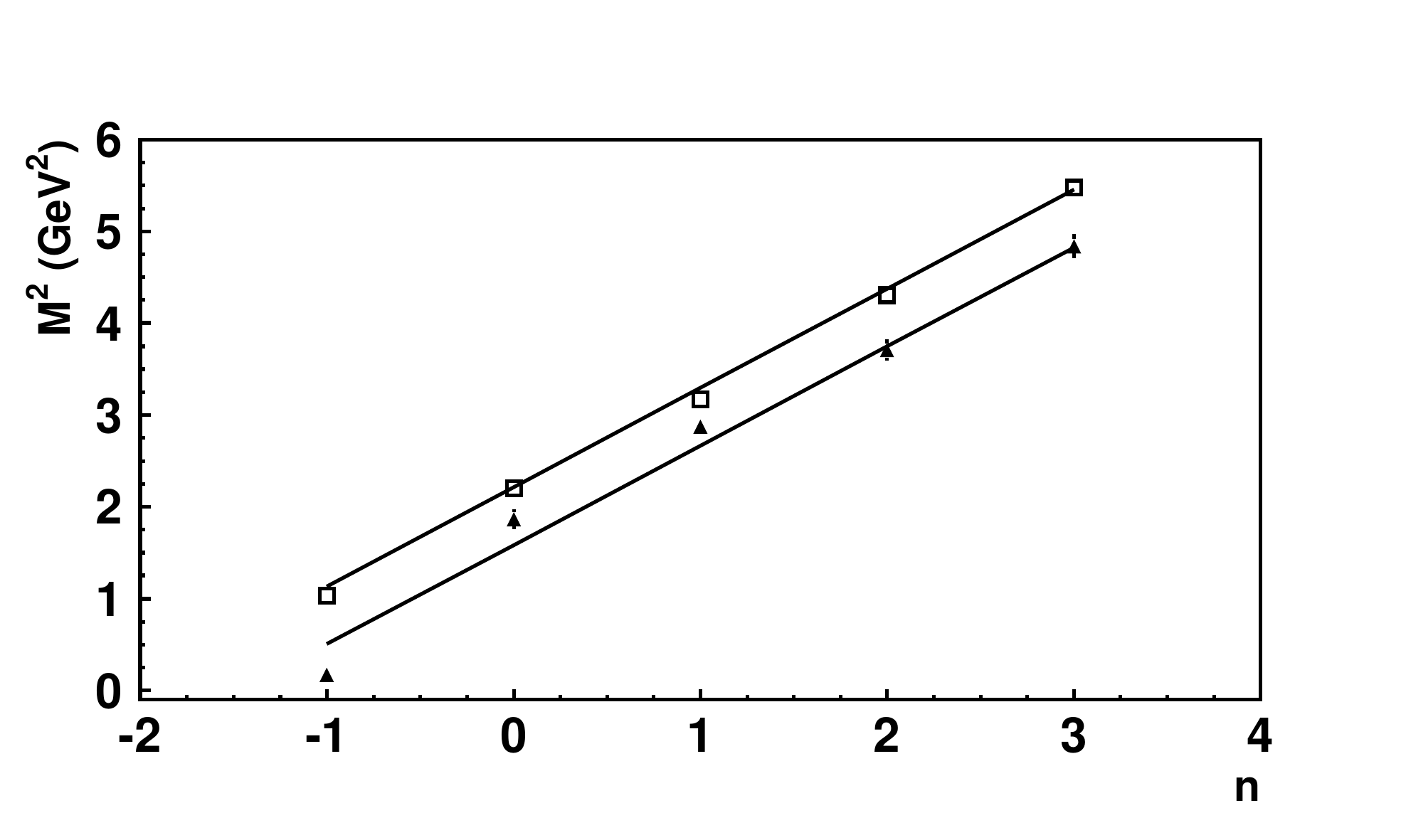}\vspace{-2mm}
\caption{\label{regge}Squared masses of mainly-octet and
mainly-singlet scalar isoscalar mesons as functions of a consecutive
number.}
\end{figure}

Figure~\ref{regge} shows $(M^2, n)$ trajectories for ``mainly-octet'' and ``mainly-singlet'' resonances. The
agreement is not perfect but astonishing for a two-parameter prediction. The  interpretation
neglects singlet-octet mixing; there could be tetraquark, meson-meson or glueball components that
can depend on $n$; final-state interactions are neglected; close-by states with the same decay
modes repel each other. There is certainly a sufficient number of reasons that may distort the
scalar-meson mass spectrum. In spite of this, the mass of none of the observed states is
incompatible with the linear trajectory by more than its half-width.

\begin{table*}[pt]
\renewcommand{\arraystretch}{1.5}
\caption{\label{MW}Pole masses and widths (in MeV) of scalar mesons. The RPP values are listed
as small numbers for comparison.\vspace{0mm}}
\centering
\begin{tabular}{cccccc}
\hline\hline
Name       & $f_0(500)$     &$f_0(1370)$     &$f_0(1710)$  &$f_0(2020)$                 &$f_0(2200)$     \\\hline
$M$  [MeV]        &  410\er 20    & 1370\er 40      &1700\er 18    & 1925\er 25                  & 2200\er 25  \\[-1.5ex]
              &\scriptsize 400\tz 550 &\scriptsize 1200\tz 1500&\scriptsize 1704\er12&\scriptsize 1992\er 16              &\scriptsize 2187\er 14\\
$\Gamma$ [GeV]  & 480\er30       &  390\er 40       & 255\er 25    & 320\er 35                   & 150\er 30\\[-1.5ex]
               &\scriptsize 400\tz 700 & \scriptsize 100\tz 500  &\scriptsize 123\er 18 & \scriptsize 442\er60               &\scriptsize $\sim 200$ \\
\hline\hline
Name       & $f_0(980)$     &$f_0(1500)$     &$f_0(1770)$  &$f_0(2100)$                 &$f_0(2330)$     \\\hline
$M$  [GeV]       &1014\er 8      &  1483\,\er\,15    &1765\er 15   &2075\er 20                   &2340\er 20\\[-1.5ex]
               &\scriptsize 990\er 20   & \scriptsize 1506\,\er\,6&                  &\scriptsize 2086$^{+20}_{-24}$&\scriptsize$\sim$2330\\
$\Gamma$ [MeV]&71\er10          & 116\er 12        & 180\er 20    & 260\er 25                   & 165\er 25\\[-1.5ex]
               &\scriptsize 10\tz 100   & \scriptsize 112\er 9     &                  & \scriptsize 284$^{+60}_{-32}$ &\scriptsize 250\er 20\\
\hline\hline\vspace{-2mm}
\end{tabular}
\renewcommand{\arraystretch}{1.5}
\caption{\label{decays}$\jpsi$ radiative decay rates in $10^{-5}$ units. Small numbers represent the RPP values,
except the $4\pi$ decay modes that gives our estimates derived from~\cite{Bai:1999mm,Bugg:2009ch}. 
The RPP values and those from Refs.~\cite{Bai:1999mm,Bugg:2009ch}  are given
with small numbers and with two digits only; statistical and systematic errors are added
quadratically.  The missing intensities in parentheses are our
estimates. Ratios for $K\bar K$ are calculated from $K_SK_S$ by multiplication with a factor 4.
Under $f_0(1750)$ we quote results listed in RPP as decays of $f_0(1710)$, $f_0(1750)$ and
$f_0(1800)$. The RPP values should be compared to the sum of our yields for $f_0(1710)$ and
$f_0(1770)$. BES~\cite{Ablikim:2018izx} uses two scalar resonances, $f_0(1710)$ and $f_0(1790)$ and
assigns most of the $K\bar K$ intensity to $f_0(1710)$. Likewise, the yield of three states at
higher mass should be compared to the RPP values for $f_0(2100)$ or $f_0(2200)$. \vspace{0mm} } \centering
\begin{tabular}{|c|ccccccc|c|c|}
\hline\hline
 $BR_{\jpsi\to\gamma f_0\to}$ & $\gamma{\pi\pi}$&$\gamma{K\bar K}$&$\gamma{\eta\eta}$&$\gamma{\eta\eta'}$&$\gamma{\omega\phi}$&\multicolumn{2}{c|}{missing}&total&unit\\[-2.ex]
                        &&&&&&\footnotesize$\gamma{4\pi}$&$\gamma\omega\omega$&&\\\hline\hline\\[-4ex]
$ f_0(500)$& 105\er20   &5\er 5&4\er 3 &$\sim$0&$\sim$0&\multicolumn{2}{c|}{$\sim$0}&114\er21&$\cdot  10^{-5}$\\\hline
$f_0(980)$&1.3\er 0.2  &0.8\er0.3&$\sim$0&$\sim$0&$\sim$0&\multicolumn{2}{c|}{$\sim$0}&2.1\er0.4&$\cdot  10^{-5}$\\\hline
$f_0(1370)$&38\er 10  &13\er4&3.5\er1&0.9\er 0.3&$\sim$0&\multicolumn{2}{c|}{14\er5}&69\er12&
\multirow{2}{*}{$\cdot  10^{-5}$}\\[-2.ex]
                        &&\scriptsize 42\er15&&&&\scriptsize 27\er9&&&\\\hline
$ f_0(1500)$&9.0\er1.7 &3\er1&1.1\er0.4&1.2\er0.5&$\sim$0&\multicolumn{2}{c|}{33\er8}&47\er9&
\multirow{2}{*}{$\cdot  10^{-5}$}\\[-1.ex]
                        &\scriptsize 10.9\er2.4&\scriptsize$2.9$\er1.2&\scriptsize $1.7^{+0.6}_{-1.4}$&\scriptsize 6.4$^{+1.0}_{-2.2}$&&\scriptsize 36\er9&&&\\\hline
$f_0(1710)$&6\er2 &23\er8&12\er4&6.5\er2.5&1\er 1&\multicolumn{2}{c|}{7\er3}&56\er10&
\multirow{3}{*}{$\cdot  10^{-5}$}\\
$f_0(1770)$&24\er8 &60\er20&7\er 1&2.5\er1.1&22\er4&\multicolumn{2}{c|}{65\er15}&181\er26&\\[-1.ex]
\scriptsize $f_0(1750)$&\scriptsize 38\er 5    &\scriptsize $99^{+10}_{\ -6}$&\scriptsize $24^{+12}_{\ -7}$&&\scriptsize 25\er 6&\scriptsize 97\er18&\scriptsize 31\er10&&\\\hline
$f_0(2020)$&42\er 10 &55\er25&10\er10&&&\multicolumn{2}{c|}{\scriptsize (38\er13)}&145\er32&
\multirow{4}{*}{$\cdot  10^{-5}$}\\
$f_0(2100)$&20\er 8 &32\er20&18\er15&&&\multicolumn{2}{c|}{\scriptsize  (38\er13)}&108\er25&\\
$f_0(2200)$&5\er2 &5\er5&0.7\er0.4&&&\multicolumn{2}{c|}{\scriptsize  (38\er13)}&49\er17&\\[-1.ex]
\scriptsize$ f_0(2100)/f_0(2200)$&\scriptsize 62\er10 &\scriptsize 109$^{+\ 8}_{-19}$&\scriptsize 11.0$^{+6.5}_{-3.0}$&&&\scriptsize 115\er41&&&\\\hline
$f_0(2330)$&4\er2 &2.5\er0.5&1.5\er0.4&&&&&8\er3&
\multirow{2}{*}{$\cdot  10^{-5}$}\\[-1.ex]
&&\scriptsize 20\er3&&&&&&&\\
\hline\hline
\end{tabular}
\vspace{30mm}
\end{table*}

Now we comment on the $\phi\omega$ decay mode. The prominent peak in Fig.~\ref{jpsi}d is ascribed
to $f_0(1770)\to\phi\omega$ decays. The BESIII collaboration interpreted the reaction as 
doubly OZI suppressed
decay~\cite{Ablikim:2012ft}. We assume that all scalar mesons have a tetraquark component as
suggested by Jaffe \cite{Jaffe:1976ig} for the light scalar meson-nonet: the price in energy to
excite a $q\bar q$ pair to orbital angular momentum $L=1$ (to $^3P_0$) is similar to the energy
required to create a new $q\bar q$ pair with all four quarks in the $S$-state. Thus, a tetraquark
component in scalar mesons should not be surprising. The tetraquark component may decay to two mesons by
rearrangement of color; thus a small tetraquark component could have a significant impact on the
decays. 
The $\phi$ and $\omega$ are orthogonal in SU(3), thus $f_0(1770)$ must have a large octet
component. The $\phi\omega$ decay is assigned to the $\frac{1}{\sqrt 6}(2u\bar ud\bar d - u\bar
us\bar s - d\bar ds\bar s)$ component that can easily disintegrate into $\phi\omega$. Scalar SU(3)
singlet mesons might have a $\frac{1}{\sqrt 3}(u\bar ud\bar d+ u\bar us\bar s+d\bar d s\bar s)$
component but their coupling to $\phi\omega$ is small since these two mesons cannot come from a
pure SU(3) singlet state.
The $\phi$ and $\omega$ are orthogonal in SU(3), thus $f_0(1770)$ must have a large octet
component. The $\phi\omega$ decay is assigned to the $\frac{1}{\sqrt 6}(2u\bar ud\bar d - u\bar
us\bar s - d\bar ds\bar s)$ component that can easily disintegrate into $\phi\omega$. Scalar SU(3)
singlet mesons might have a $\frac{1}{\sqrt 3}(u\bar ud\bar d+ u\bar us\bar s+d\bar d s\bar s)$
component but their coupling to $\phi\omega$ is small since these two mesons cannot come from a
pure SU(3) singlet state.

\section{Multiparticle decays}
Scalar mesons may also decay into multi-meson final states. This fraction is determined here as
missing intensity in the mass range where data on $\pi\pi$ elastic scattering are available. The
results are also given in Table~\ref{decays} and compared to earlier determinations. The reaction
$J/\psi\to \gamma \pi^+\pi^-\pi^+\pi^-$ has been studied in Refs.~\cite{Bai:1999mm,Bugg:2009ch}.
The partial wave analyses determined $\sigma\sigma$ as main decay mode of the scalar mesons. Then,
the yields seen in $2\pi^+ 2\pi^-$ need to be multiplied by 9/4 to get the full four-pion yield.
These estimated yields for $J/\psi\to \gamma 4\pi$ are given in Table~\ref{decays} by small
numbers.  Assuming different decay modes (like those reported in Table III in \cite{Abele:2001pv})
leads to small changes only in the four-pion yields. The $\omega\omega$ yield, determined in
$J/\psi\to \gamma \omega\omega$~\cite{Ablikim:2006ca}, is unexpectedly large and inconsistent with
the small $\rho^0\rho^0$ yield. 

The missing intensity of $f_0(1370)$ reported here is not inconsistent with the branching ratio found in 
radiative decays $J/\psi\to \gamma \pi^+\pi^-\pi^+\pi^-$ but contradicts the findings from 
$\bar pN$ annihilation into five pions and from central production of four pions. This discrepancy
can only be resolved by analyzing data on $J/\psi\to \gamma 4\pi$ 
and  $\bar pN$ annihilation in a coupled-channel analysis. The inclusion of
both data sets seems to be of particular importance.

First analyses of the reactions $J/\psi\to \gamma \rho\rho$ and 
$J/\psi\to \gamma \omega\omega$ \cite{Baltrusaitis:1985zi,Baltrusaitis:1985nd}
revealed only small scalar contributions.
A few scalar resonances found here were identified in $J/\psi\to \gamma
4\pi$~\cite{Bai:1999mm,Bugg:2009ch} and $J/\psi\to \gamma \omega\omega$~\cite{Ablikim:2006ca}.
We compare our missing intensities for $f_0(1710)$ and $f_0(1770)$ with the measured
$4\pi$ and the $\omega\omega$ decay modes assigned to $f_0(1750)$. Our
missing intensities are mostly well compatible with the measured $4\pi$ and $2\omega$ yields. 
For the high-mass
states we distribute the measured $4\pi$ intensity equally among the
three resonances with masses close to the $f_0(2100)$. Our estimated intensities are 
given in Table~\ref{decays}
in parentheses. Changing how the $4\pi$ intensity is distributed has little effect on the properties
of the peak shown in Fig.~\ref{fig}.

\section{The integrated yield}

\begin{figure*}[pt]
\begin{center}
\begin{tabular}{ll}
\begin{overpic}[scale=0.44]{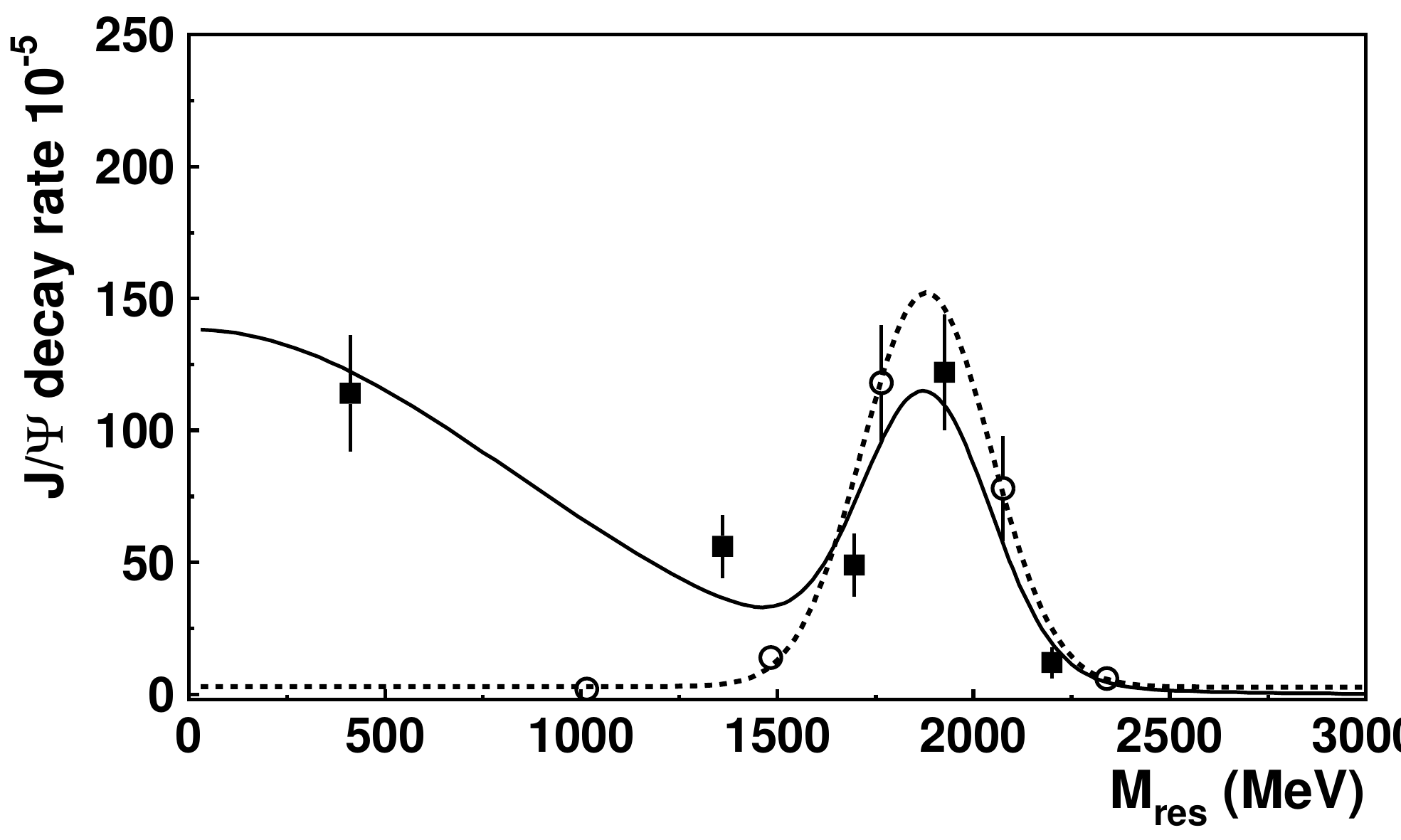}
\put(90,50){\huge a}
\end{overpic}&
\begin{overpic}[scale=0.44]{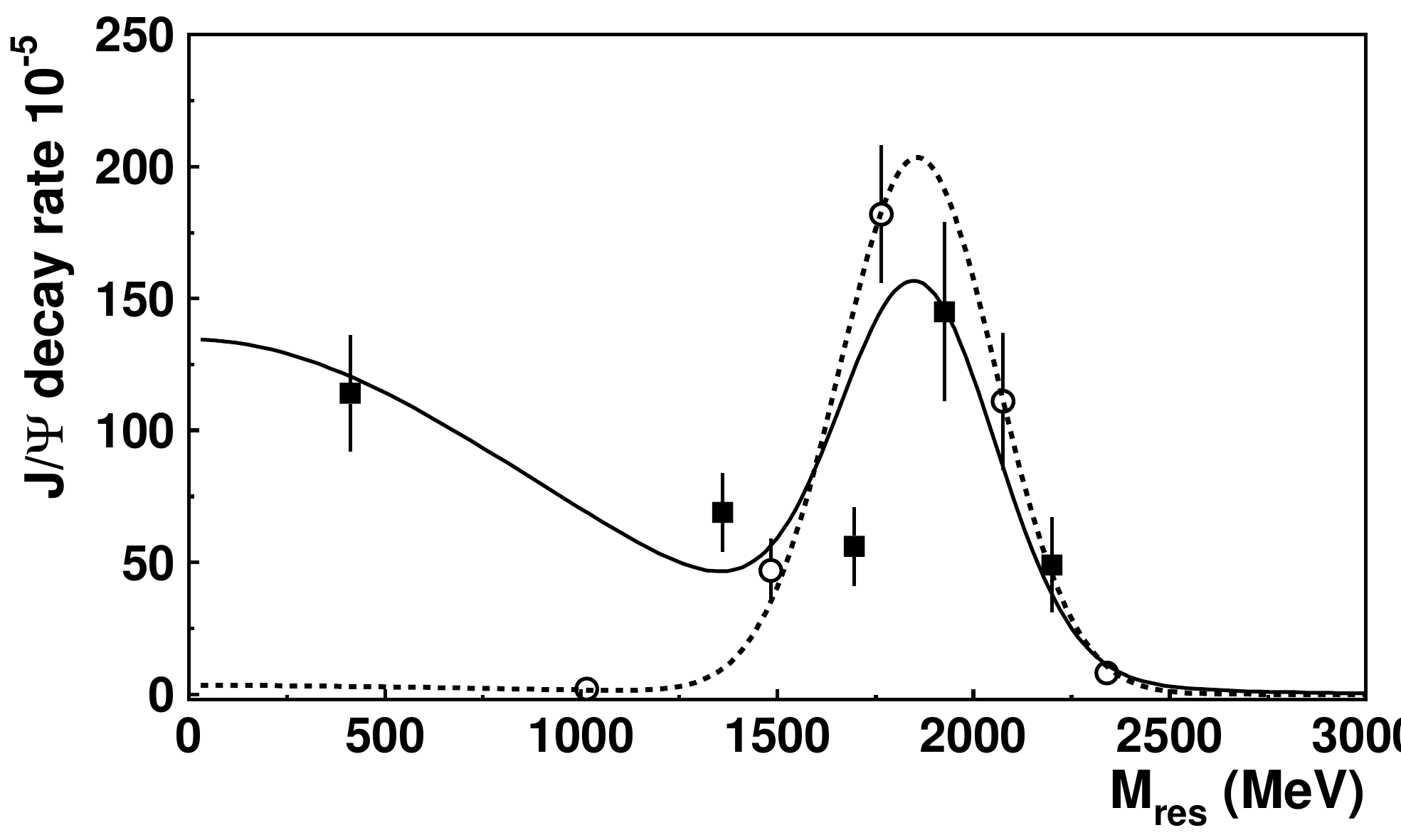}
\put(90,50){\huge b}
\end{overpic}
\end{tabular}\vspace{-3mm}
\end{center}
\caption{\label{fig}Yield of radiatively produced
scalar isoscalar octet mesons (open circles) and singlet (full
squares) mesons. a) Yield for $\pi\pi$, $K\bar K$, $\eta\eta$,
$\eta\eta'$, and $\phi\omega$ decays. b) Yield when $4\pi$ decays
and $\omega\omega$ are included.\vspace{5mm}
 }
\end{figure*}

Figure~\ref{fig} shows the yields of scalar SU(3)-singlet and octet resonances  as functions of their mass,
for two-body decays in Fig.~\ref{fig}a and for all decay modes (except $6\pi$) in Fig.~\ref{fig}b.
SU(3)-octet mesons are produced in a limited mass range only. In this mass range, a clear peak shows up. 
SU(3)-singlet mesons are produced over the full mass range but at about 1900\,MeV, their yield is enhanced.
Obviously, the two gluons from radiative $J/\psi$ decays couple to SU(3) singlet mesons in the full
mass range while octet mesons are formed only in a very limited mass range. But both,
octet and singlet scalar isoscalar mesons are formed preferentially in the 1700 to 2100\,MeV mass range.

The peak structure is unlikely to be explained as kinematics effect. Billoire {\it et al.}
\cite{Billoire:1978xt} have calculated the mass spectrum of two gluons produced in radiative
$J/\psi$ decay. For scalar quantum numbers, the distribution has a maximum at about 2100\,MeV and
goes down smoothly in both directions. K\"orner {\it et al.} \cite{Korner:1982vg} calculated the
(squared) amplitude to produce scalar mesons in radiative $J/\psi$ decays. The smooth amplitude
does not show any peak structure, neither.

\section{The scalar glueball}
We suggest to interpret this enhancement as the scalar glueball of lowest mass. The scalar
isoscalar mesons that we assigned to the SU(3) octet seem to be produced only via their mixing with
the glueball. Indeed, $J/\psi\to\gamma f_0^8$ decays are expected to be suppressed: two gluons
cannot couple to one SU(3) octet meson. Mesons interpreted as singlet scalar isoscalar mesons are
produced over the full mass range. This finding supports strongly the interpretation of the scalar
mesons as belonging to SU(3) singlet and octet.

Figure~\ref{fig}a and \ref{fig}b are fitted with a Breit-Wigner amplitude. We determined the 
yield of the scalar glueball as sum of the yield of ``octet" scalar mesons
plus the yield of ``singlet" scalar mesons above a suitably chosen phenomenological
background.  Different background shapes were assumed. In the shown one, a background 
of the form $x\cdot\exp{\{-\alpha M^2\}}$ ($x=149, \alpha=0.73$/GeV$^2$) was used. 
For Fig.~\ref{fig}a, we find 
$(M, \Gamma) = (1872, 332)$\,MeV, for Fig.~\ref{fig}b $(M, \Gamma) = (1856, 396)$\,MeV.
The results depend on the background chosen. 
From the spread of results when the background function is changed, we estimate the uncertainty. 
Our best estimate for the scalar glueball mass and width is given as
\be
\hspace{-5mm}M_G=(1865\pm 25^{\,+10}_{\,-30})\,{\rm MeV}\quad \Gamma_G= (370\pm 
50^{\,+30}_{\,-20})\,{\rm MeV}\nonumber.
\ee
The integrated yields depend on the not well-known $4\pi$ and $\omega\omega$
(and unknown $6\pi$) contributions. The (observed) yield of octet scalar isoscalar mesons
plus the yield of singlet mesons above the background is determined to
\be
Y_{J/\psi\to\gamma G}=(5.8\pm 1.0)\,10^{-3}.\nonumber
\ee
The two states with the largest glueball component are $f_0(1770)$ and $f_0(2020)$. The
``mainly octet'' $f_0(1770)$ acquires a glueball component (and is no longer ``mainly octet'', only
the $q\bar q$ and tetraquarks components belong to the octet). We suggest their wave functions
could contain a small $q\bar q$ component (a $q\bar q$ {\it seed}), a small tetraquark component as
discussed above, and a large glueball component. At the present statistical level,
there seem to be no direct decays of the glueball to mesons; the
two gluons forming a glueball are seen only since the glueball
mixes with scalar mesons. We observe no ``extra'' state.

\section{Summary}

Summarizing, we have performed a first coupled-channel analysis of the $S$-wave partial-wave
amplitudes for $J/\psi$ radiative decays into $\pi\pi$, $K_SK_S$, $\eta\eta$, and $\phi\omega$
decays. The fits were constrained by a large number of further data. The observed pattern of peaks
and valleys in the $\pi\pi$ and $K \bar K$ invariant mass distributions depends critically on the
interference between neighboring states. We are convinced that only a coupled-channel analysis has
the sensitivity to identify reliably the position of resonances.

Scalar mesons seem to show up as mainly-singlet and mainly-octet states in SU(3). The masses of
both, of singlet and octet states, are compatible with a linear $(M^2, n)$ behavior. 
Only the $f_0(500)$, mostly interpreted as dynamically generated $\pi\pi$ molecule,
does not fall onto the trajectory. The $\omega\phi$ decay mode of some scalar resonances suggests
that these may have a tetraquark component as it was suggested for the lowest-mass scalar-meson
nonet by Jaffe  45 years ago. Thus, a simple picture of the scalar-meson mass spectrum has emerged.
The yield of scalar mesons in radiative $J/\psi$ decays shows a significant structure that we
propose to interpret as scalar glueball.

The BESIII collaboration has recorded data with significantly improved quality and statistics.
It seems very important to repeat this analysis with the full statistics and
including all final states into which scalar mesons can decay.\\

\section*{Acknowledgement}
Funded by the NSFC and the Deutsche Forschungsgemeinschaft (DFG, German 
Research Foundation) through the funds provided to the Sino-German Collaborative
Research Center TRR110 “Symmetries and the Emergence of Structure in QCD”
(NSFC Grant No. 12070131001, DFG Project-ID 196253076 - TRR 110) and 
the Russian Science Foundation (RSF 16-12-10267).

\end{document}